\documentclass[sigconf]{acmart}
\usepackage{color}
\usepackage{comment}
\usepackage{algorithm}
\usepackage{algorithmic}

\AtBeginDocument{%
  \providecommand\BibTeX{{%
    \normalfont B\kern-0.5em{\scshape i\kern-0.25em b}\kern-0.8em\TeX}}}

\copyrightyear{2021}
\acmYear{2021}
\setcopyright{acmcopyright}
\acmDOI{10.1145/1122445.1122456}

\acmConference[CIKM '21] {Proceedings of the 30th ACM International Conference on Information and Knowledge Management}{November 1--5, 2021}{Virtual Event, Australia.}
\acmBooktitle{Proceedings of the 30th ACM Int'l Conf. on Information and Knowledge Management (CIKM '21), November 1--5, 2021, Virtual Event, Australia}
\acmPrice{15.00}
\acmISBN{978-1-4503-8446-9/21/11}
\acmDOI{10.1145/3459637.3482241}

\begin{document}
\fancyhead{}

\title{CMML: Contextual Modulation Meta Learning \\ for Cold-Start Recommendation}

\author{Xidong Feng$^{1,*}$, Chen Chen$^{2,*}$, Dong Li$^{2}$, Mengchen Zhao$^{2}$, Jianye Hao$^{2}$, Jun Wang$^{1, \#}$}\thanks{$*$ Equal Contribution, Email: xidong.feng.20@ucl.ac.uk, chenchen9@huawei.com\\
$\#$ Corresponding Author. Email: junwang@cs.ucl.ac.uk
}

\affiliation{
\institution{$^1$ University College London $^2$ Noah's Ark Lab, Huawei}
\city{}
\country{}
}

\renewcommand{\shortauthors}{Trovato and Tobin, et al.}

\begin{abstract}

Practical recommender systems experience a cold-start problem when observed user-item interactions in the history are insufficient. Meta learning, especially gradient based one,  can be adopted to tackle this problem by learning initial parameters of the model and thus allowing fast adaptation to a specific task from limited data examples. Though with significant performance improvement, it commonly suffers from two critical issues: the non-compatibility with mainstream industrial deployment and the heavy computational burdens, both due to the inner-loop gradient operation. These two issues make them hard to be applied in practical recommender systems. To enjoy the benefits of meta learning framework and mitigate these problems, we propose a recommendation framework called \textbf{C}ontextual \textbf{M}odulation \textbf{M}eta \textbf{L}earning (CMML). CMML is composed of fully feed-forward operations so it is computationally efficient and completely compatible with the mainstream industrial deployment. It consists of, a context encoder that can generate context embedding to represent a specific task, a hybrid context generator that aggregates specific user-item features with task-level context, and a contextual modulation network, which can modulate the recommendation model to adapt effectively. We validate our approach on both scenario-specific and user-specific cold-start setting on various real-world datasets, showing CMML can achieve comparable or even better performance with gradient based methods yet with  higher computational efficiency and better interpretability.



\end{abstract}

\ccsdesc[500]{Information systems~Information retrieval}

\keywords{Recommender system; Cold-start problem; Meta learning}


\maketitle
\section{Introduction}
Personalized recommendation has been rapidly developed and thoroughly influenced various fields including web services, E-commerce, and social media. Based on user's feature and past interaction history with items, the personalized recommender system can generate reasonable recommendation results for user's preferences. Among the personalized recommendation approaches, the data-driven deep learning based recommendation algorithms have gained increasing attention because of their superior performance compared with traditional recommendation algorithms.
However, both traditional and data-driven machine learning based recommendation algorithms struggle to tackle the cold-start problem, since the recommender system can only get access to very limited user's interaction history. To address cold-start recommendation problems, some works directly utilize content information such as user's profiles \cite{adomavicius2011context,van2013deep} or incorporate it into traditional collaborative filtering \cite{wei2017collaborative, cheng2016wide, strub2016hybrid},  but they still often fail to generalize to users with only a few interactions. Some other works \cite{man2017cross, hu2018conet} try to solve the cold-start problem by transferring domain knowledge from cross-domain datasets, but they still cannot get rid of the need for shared examples from different domains.\par
Meta learning,  also known as leaning to learn, is a prominent machine learning paradigm aiming at learning meta knowledge among tasks to achieve fast adaptation with limited data examples when facing a new task. Inspired by this, some works \cite{lee2019melu,du2019sequential, dong2020mamo, wei2020fast} leverage the latest progress on  meta learning to model and solve the cold-start recommendation problem. They treat different entities (like users, scenarios) in recommender system as tasks, so the cold-start recommendation problem can be transformed into a new-task adaptation problem in meta learning. They commonly adopt MAML \cite{finn2017model}, which is a representative gradient based meta learning framework to solve the problem. Specifically, they try to optimize the recommendation model's initial parameters, to make it capable of fast adaptation from limited data examples through taking only a few gradient steps. Practically, they have achieved  significant improvements over previous deep learning based recommendation algorithms, which validates the effectiveness of introducing meta learning framework into the cold-start problem.\par
However, they have also brought some new issues which can be basically summarized into two parts. (1) The additional inner-loop gradient adaptation will lead to much lower computational efficiency with respect to both the training phase and inference phase. (2) The inner-loop gradient operator is incompatible with current industrial recommender system's framework, where most recommendation algorithms only consist of feed-forward network like Multilayer perception (MLP). These two problems  prevent the broad use of these gradient based approach on cold-start problems. \par
In order to both enjoy the benefits from meta learning and mitigate the computational or deployment problems, we are trying to answer the following question: \textbf{from limited data examples, how can we conduct meta learning model adaptation based on feed-forward operations, rather than backward propagation as done in the gradient based approach?} An intuitive idea is to directly map limited samples to continuous representation for guiding the model adaptation. Specifically, we need a powerful embedding network for information extraction from samples and a flexible modulation network for model's adaptation. These modules can be trained via the meta learning framework and thus equip the model with fast adaptation ability even given limited data.\par
Based on these motivations, we propose a \textbf{C}ontextual \textbf{M}odulation \textbf{M}eta \textbf{L}earning (CMML)  recommendation framework for cold-start problems. Note that the term 
\textit{contextual} here does not represent any specific context information, such as time, user's features. Here we treat the interaction history as contextual information to reveal entities' (users/scenarios) implicit features. In our framework, instead of utilizing inner-loop gradient adaptation, we make use of modulation techniques for model adaptation when a new task comes. Specifically, our model consists of three parts: (1) A context encoder that maps the limited data examples to effective context aggregation. 
(2) A hybrid context generator, which combines the above encoded task-level context (scenario/user-specific information) and instance-level feature (user-item features) together. By modelling the interaction between task-level context and local user-item pair's features, flexible hybrid context can largely enhance the model's capacity.    (3) A context based modulation network, which can modulate the neural network for fast adaptation based on context generated by two parts above. Compared with previous gradient based meta recommendation algorithms, one of the most appealing  characteristics for our method is that it is fully composed of feed-forward neural network and  successfully gets rid of the gradient adaptation. It can significantly improves the computational efficiency  and  is compatible with industrial recommender systems.
\par
The remainder of this paper is organized as follows: in Section \ref{preliminary}, we introduce the problem formulation for meta learning based cold-start recommendation and the setting for two representative types of important cold-start problems. Section \ref{method} depicts the whole framework of the proposed CMML approach and also provides computational efficiency analysis. Experimental results about detailed experimental settings, comparison, visualization, ablation study are shown in Section \ref{experiment}. We review related work in Section \ref{related work}. Finally, we conclude our paper in Section \ref{conclusion}.
\section{PRELIMINARY}
\label{preliminary}
In this section, we formally define the problem formulation for meta learning based recommendation and detail the settings for scenario-specific and user-specific meta recommendation problems.
\subsection{Problem Formulation}
Traditional recommendation is composed of two sets, user set $\mathcal{U}$, item set $\mathcal{I}$, which includes user's features and item's features, respectively. It conducts the mapping function: $\mathcal{U} \times \mathcal{I} \rightarrow \mathbb{R}$, which means they will generate prediction results based on users' and items' features. For context-aware recommendation problem, the recommender system is required to conduct the mapping function: $\mathcal{U} \times \mathcal{I} \times \mathcal{C} \rightarrow \mathbb{R}$, where $C$ represents contextual information set. In this paper, we are trying to solve cold-start problem by meta learning based recommendation, where model can conduct task adaptation based on limited data from specific entities (users/scenarios). For meta learning based  recommendation, the framework needs to transform interaction history into implicit contextual information and conduct rapid adaptation to predict results. This mapping function corresponds to $\mathcal{U} \times \mathcal{I} \times \{\mathcal{U}, \mathcal{I}\}_h \rightarrow \mathcal{U} \times \mathcal{I} \times \mathcal{C} \rightarrow \mathbb{R}$, where $\{\mathcal{U}, \mathcal{I}\}_h$ represents interaction history. The mapping function  $\{\mathcal{U}, \mathcal{I}\}_h \rightarrow \mathcal{C}$ can be gradient descent operator like \cite{lee2019melu}, \cite{du2019sequential}, or the contextual modulation framework introduced in this paper.\par
For meta learning problems, there are some important terms that need to clarify for clear comprehension. The terms meta-training and meta-testing: these two terms correspond to the standard definition of training set and testing set in traditional machine learning. The major difference is that data examples in meta learning are tasks.
Support-set and query-set: in meta learning, it commonly assumes that there exist limited labeled data examples from which the model can take rapid adaptation. Such labeled samples are denoted as support-set. The samples which need to be predicted are denoted as query-set. Note that in the meta-training phase, we will get labels for both support-set and query-set. The labeled  support-set is used for inner-loop adaptation, while the labeled query-set is used for outer-loop  optimization. In the meta-testing phase, we will only get labeled data examples from support-set, and leaving the  adapted model evaluated on the unlabeled query-set.\par
With above definitions, the meta recommendation problems can be formulated as the following machine learning problem. It assumes that the recommendation tasks $T$ subjects to a fixed task distribution $P(\tau)$. The final objective of model $f_{\Theta}$ is to minimize the expectation loss on the prediction results for query-set samples $D_{T}^{\text {query}}$ ($D_{T}^q$) after conducting adaptation from  support-set samples $D_{T}^{\text {support}}$($D_{T}^s$). Thus, the objective function can be written as follows:
\begin{equation}
    \min _{\Theta} \mathbb{E}_{T\sim P(\tau)}\left[\mathcal{L}_{\Theta}\left(D_{T}^{\text {query }} \mid D_{T}^{\text {support}}\right)\right],
\end{equation}
where we denote all parameters and loss function with $\Theta$ and $\mathcal{L}$ respectively. For meta recommendation problem, the $D_{T}^{\text {support}}$ here represents interaction history for recommended entities and $D_{T}^{\text {query}}$ denotes the user-item pairs the adapted model needs to predict.\par
For our proposed CMML framework, we split the parameters $\Theta$ into two parts: backbone network's parameters $\Phi$ and meta model parameters $\Theta_M$.
Practically, we split the tasks of the dataset into meta-training tasks $D_{mtr}$ and meta-testing tasks $D_{mte}$ and try to minimize the empirical risks on $D_{mtr}$. 
In all, the loss function can be represented as Equation (\ref{equ:cmml}) and (\ref{equ:cmml1}).
\begin{equation}
\label{equ:cmml}
     \min _{\Theta_M, \Phi} \mathbb{E}_{T\sim P(D_{mtr})}\left[\mathcal{L}_{\Theta_M, \Phi}\left(D_{T}^{\text {query }} \mid D_{T}^{\text {support}}\right)\right],
\end{equation}
\begin{equation}
 \mathcal{L}_{\Theta_M, \Phi}\left(D_{T}^{\text {q}} \mid D_{T}^{\text {s}}\right)=\sum_{\left(u_{k}, i_{k}\right) \in D_{T}^{\text {q }}} \frac{\ell\left(f_{\Theta_M, \Phi}(u_{k}, i_{k} ; \{u,i\}_{h}), y_k\right)}{\left|D_{T}^{\text {q }}\right|},
 \label{equ:cmml1}
\end{equation}
where $f_{\Theta_M, \Phi}$ represents the whole model, $(u_k, i_k), y_k$ denotes the $k$-th user-item pair and corresponding label respectively. $\{u,i\}_h$ is the user-item interaction history in support-set. Note that we neglect the symbol of real labels for $\{u,i\}_h$ in support-set in the whole paper for notation simplicity.
\subsection{Scenario/User-specific Settings}
In this paper, we choose two cold-start recommendation settings, scenario-specific \cite{du2019sequential} and user-specific settings \cite{lee2019melu}, to show the effectiveness and generalization ability of our proposed framework.
Note that our framework is not limited to these two settings and can also be applied into broader industrial recommendation problems.\par
The scenario-specific recommendation utilizes scenario characteristics to enhance the recommendation performance. For instance, for sports related scenarios, items related with outdoor activities should gain more popularity. In meta learning, our framework treats different scenarios as different tasks. Without explicit information, the algorithm needs to extract implicit scenario characteristics from limited support-set, and conducts scenario-specific adpatation.
Following \cite{du2019sequential}, we model the learning objective as a click-through-rate(CTR) prediction problem and utilize hinge loss on query-set as our meta objective function. We denote $i^{+}$ and $i^{-}$ as positive items and negative items for CTR. Following  Equation (\ref{equ:cmml1}), the loss function can be written as follows:
\begin{equation}
\ell=\max \left(0, 1-f_{\Phi,\Theta_M}\left(u, i^{+} ; \{u,i\}_h\right)+f_{\Phi,\Theta_M}\left(u, i^{-};\{u,i\}_h\right)\right).
\end{equation}
\par
For user-specific meta recommendation setting, the tasks in meta learning framework corresponding to users $U$ in recommendation. The interaction history of $\{u, i\}_h$ will be treated as the support-set, which can implicitly reveal the user's profiles and preferences. We consider a score regression problem \cite{lee2019melu} in this setting. The objective of meta model is to minimize the mean square error on new user-item pairs' prediction scores. Following Equation (\ref{equ:cmml1}), the loss function can be written as follows:
\begin{equation}
\ell=(y_{u,i}-f_{\Phi, \Theta_M}\left(u, i;\{u,i\}_h\right))^2,
\end{equation}
where $y_{u,i}$ is the real score for user-item pair $(u,i)$.

\section{CMML}
\label{method}
In this part, we illustrate the backbone network and detailed structure for our proposed CMML framework. In all, the framework consists of three components: a context encoder that can aggregate the interaction history in the support-set for implicit contextual information extraction and effective task representation, a hybrid context generator that aggregates task-level information and instance-level user-item features, and modulation network, which modulates the backbone network for fast adaptation on new scenarios/users.\par
\begin{figure*}[t]
    \centering
    \includegraphics[width=0.8\linewidth]{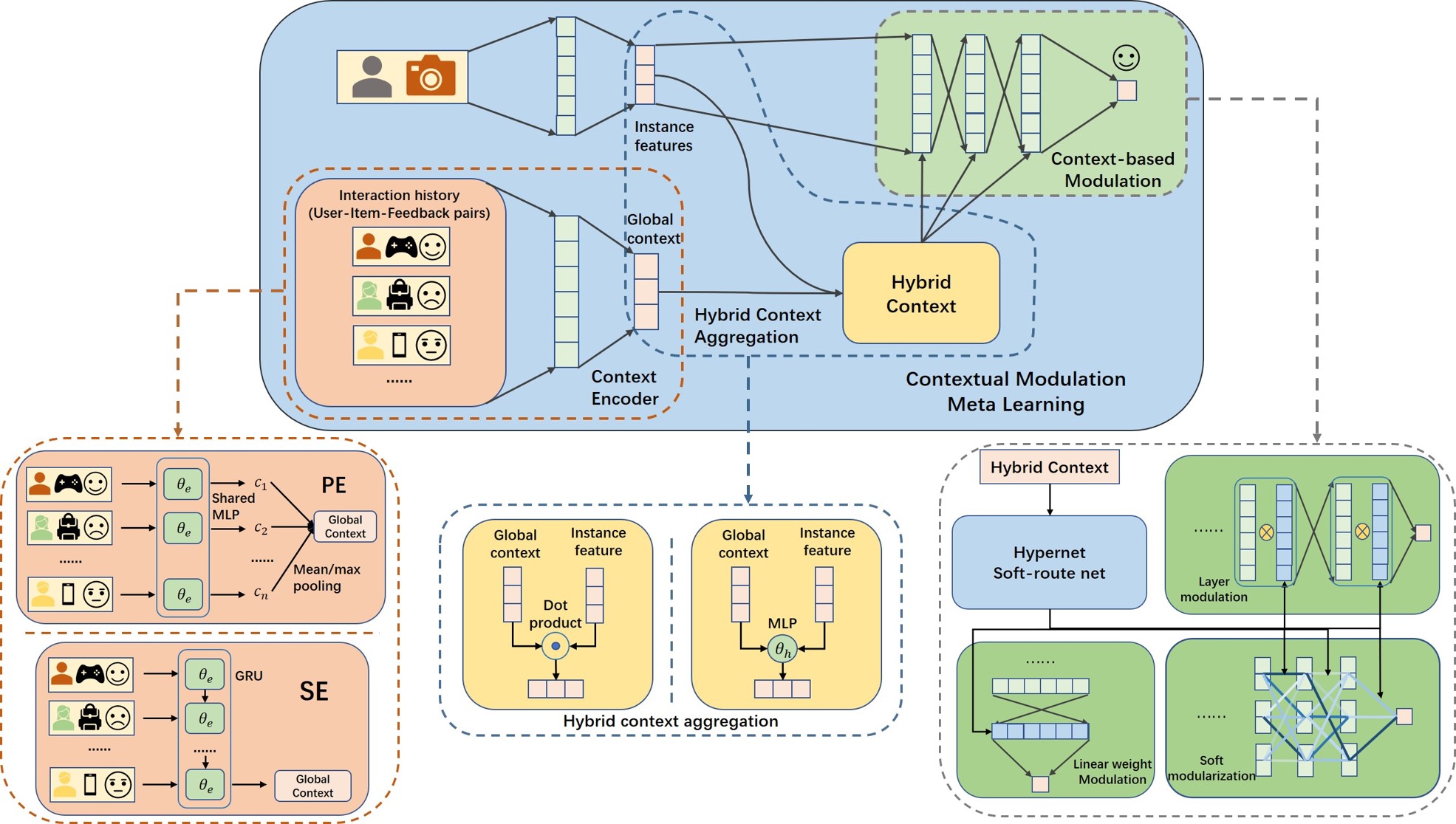}
    \caption{CMML framework}
    \label{fig:framework}
\end{figure*}

The detailed procedure is summarized as follows. First, the context encoder will map interaction history for task-level contextual information $C$. Second, the embedded context will be combined with specific user-item feature embedding, generating hybrid context. Conditioned on hybrid context, the backbone network will be modulated  for new scenario/user adaptation. The detailed algorithm chart is shown in Algorithm \ref{alg:algorithm-context}. We also provide Figure \ref{fig:framework} for better illustration of the overall framework and specific modules.
\begin{algorithm}
    \caption{Contextual modulation meta learning}
    \label{alg:algorithm-context}
    \begin{algorithmic}[1]
    \REQUIRE the meta-training dataset $D_{mtr}$
    \FOR{Epoch e in training epochs}
    \STATE Sample a batch of users/scenarios from $D_{mtr}$ as $D_{mb}$
    \STATE $\mathcal{L}_{meta}=0$
    \FOR{each user/scenario T in $D_{mb}$}
    \STATE Obtain user-item pairs in support-set $D_T^s$
    \STATE From support-set $D_T^s$, global context $C$ is obtained by context encoder using equ \ref{equ:pooling}, \ref{equ:gru}
    \STATE For each user-item pair in query-set $D_T^q$, conduct hybrid context aggregation by equ \ref{equ:hybrid}
    \STATE Obtain contextual modulation model by equ \ref{equ: liner modulation}, \ref{equ:film}, \ref{equ:base}
    \STATE Evaluate the loss function $l_T$ on query-set's labels, $\mathcal{L}_{meta} \leftarrow \mathcal{L}_{meta} +  \frac{l_T}{|D_{mb}|}$
    \ENDFOR 
    \STATE $\Theta_{meta} \leftarrow \Theta_{M} - \alpha \nabla_{\Theta_{M}} \mathcal{L}_{meta}$
    \STATE $\Phi \leftarrow \Phi- \alpha \nabla_{\Phi} \mathcal{L}_{meta}$
    \ENDFOR
    \end{algorithmic}
\end{algorithm}
\subsection{Backbone Network Structure}
The recommender backbone network maps user-item features to prediction results, and will be modulated by the modulation modules introduced in Section \ref{3.4}. We adopt a common feed-forward neural network as the backbone network structure for simplicity. But note that our framework can also be plugged into many sophisticated backbone networks. The whole backbone network consists of three components: embedding layers, hidden layers, and output layer. All parameters mentioned here belong to $\Phi$.\par
The embedding layer consists of user embedding matrices $M_u$ and item embedding matrices $M_i$. The high-dimensional user features/ids $x_u$ and item features/ids $x_i$ can be transformed into low dimensional representations $e_u, e_i$ by $e_u = M_u x_u$ and $e_i = M_i x_i$. Here $M_u, M_i$  can be learned through the end to end training process \cite{lee2019melu} or be pre-generated \cite{du2019sequential} by collaborative filtering \cite{koren2009matrix}. The hidden layers are ReLU activated MLP which map the concatenated user-item feature embedding $(e_u, e_i)$ to the continuous representation $h_{ui}$. The final output is a linear layer $f_o$, which maps the hidden layer's output to final score $o_{ui}$ for specific user-item pair. The transformation corresponds to $o_{ui}=w^T h_{ui} + b$, where $w, b$ are weights and bias for the last layer respectively.
\subsection{Context Encoder}
Context encoder can be regarded as the embedding network that maps interaction history of user-item pairs to implicit contextual information: $\{\mathcal{U}, \mathcal{I}\}_h \rightarrow \mathcal{C}$. Here we offer two alternatives: pooling aggregated encoder (PE) and sequential aggregated encoder (SE).
\subsubsection{Pooling aggregated encoder}
For pooling aggregated encoder, it maps $n$ user-item pairs $\{u_k, i_k\}_{k=1}^n\in D_T^s$ to user/scenario context. Each user-item pair $(u,i)$ will be firstly mapped to the corresponding embedding ${e_u, e_i}$ by embedding layers of the backbone network. Then the embedding is fed into ReLU activated MLP $f_{\theta_e}$, which is shared among all user-item pairs. At last, mean/max pooling operation will be conducted to aggregate user-item pairs to single user/scenario context $C$. The above process can be formulated as Equation (\ref{equ:pooling}):
\begin{equation}
\label{equ:pooling}
\begin{aligned}
    C_k &= f_{\theta_e}(\{M_u x_u, M_i x_i\}_k),\\
     C &= \text{Mean/Max Pooling}(\{C_k\}_{n}),
\end{aligned}    
\end{equation}
where $f^c_{\theta_e}$ denotes the context encoder $f^c$ parameterized by meta parameters $\theta_e \in \Theta_M$, $C_k$ is the context embedding for $k$-th user-item pair and $C$ is the final task-specific context. \par
Pooling aggregated encoder is a fairly simple model for implementation and it has the property of permutational invariance, which means the output of neural network will not be affected  for different permutations of inputs. Since we usually get no access to explicit sequential data in our setting, the permutation for user-item pairs contains no useful information for identifying task-specific features. The permutational invariance property enables the model to neglect the sequential order for items in support-set.
\subsubsection{Sequential aggregated encoder}
To further enhance the information aggregation ability of the context encoder, we also provide  sequential aggregated encoder (SE).  We treat all user-item pairs $\{u_k, i_k\}_{k=1}^n\in D_s^T$ as a sequence of information and leverage sequential model like Gated Recurrent Unit(GRU) \cite{cho2014learning} to handle the representation of the support-set. The user-item pairs are firstly fed into GRU network sequentially, and the embedded sequence representation from it will be mapped into the global context by ReLU activated MLP. The process can be formulated as Equation (\ref{equ:gru}):
\begin{equation}
\label{equ:gru}
\begin{aligned}
     \{h_k\}_{k=1}^n, \{o_k\}_{k=1}^n &=\text{GRU}_{\theta_{e}^1}(\{M_u x_u^k, M_i x_i^k\}_{k=1}^n),\\
     C &= f_{\theta_{e}^2}(o_n),
\end{aligned}
\end{equation}
where $\text{GRU}_{\theta_e^1}$ is the GRU based context encoder parameterized by meta parameters $\theta_e^1\in \Theta_M$,  $\{h_k\}_{k=1}^n, \{o_k\}_{k=1}^n$ denote the output sequence of the GRU model and $f_{\theta_{e}^2}$ represents the MLP parameterized by meta parameters $\theta_e^2 \in \Theta_M$ that maps sequence embedding $o_n$ into context $C$. \par
It deserves to be pointed that since data examples in support-set usually contain no sequential information, we choose sequential model because of its better context aggregation ability compared with mean-pooling operation in PE rather than its sequential property. Note that though in principle, the sequential aggregated encoder does not hold the property of permutational invariance, we can still randomly permutate the support-set at each iteration in practice so the sequential aggregated encoder can learn to be agnostic to the order information. This is also the technique utilized in \cite{hamilton2017inductive}.
\subsection{Hybrid Context Generator}
Note that the context encoder in previous section only extracts user/scenario information for task-level context from support-set $D_T^{\text{support}}$. However, the task-level context will be the same for different user-item pairs in the query-set. The interaction between task-level context and specific instance-level user-item feature is neglected. Thus, in this section, we provide the module of hybrid context generator to combine both information effectively.\par
There are many classical works in recommender system's literature like \cite{wang2017deep, guo2017deepfm} about how to conduct low-order and high-order feature interactions. Usually, low-order interaction is obtained by taking low-order computation like dot-product among features while high-order interaction is obtained by deep neural network. We borrow the idea of this to formulate the hybrid context generator. By taking dot-product or MLP to model the relationship of these features, we provide two ways to combine information in Equation (\ref{equ:hybrid}).
\begin{equation}
\label{equ:hybrid}
\begin{aligned}
   C^h_{u_k, i_k} &= C \odot  \{M_u x_{u_k}, M_i x_{i_k}\},\\
   C^h_{u_k, i_k} &= \text{MLP}_{\theta_d}(C, \{M_u x_{u_k}, M_i x_{i_k}\}),
\end{aligned}
\end{equation}
where $C^h_{u_k, i_k}$ denotes the specific hybrid context for specific user-item pair $\{u_k, i_k\}$, $\theta_d \in \Theta_M$  represents the  meta parameters of hybrid context generator.
\subsection{Context based Modulation Network}
\label{3.4}
After obtaining hybrid context, the next step for our framework is to effectively adapt for task-specific model. Specifically, we propose context based modulation network which modulates the backbone network $\Phi$ for fast adaptation. In this section, we provide three ways to conduct network modulation: weight modulation, layer modulation and soft modularization. And we will  use $C_h$ to denote the  specific hybrid context $C^{h}_{u_k,i_k}$ in  Equation (\ref{equ:hybrid})  for simplicity in the following.
\subsubsection{\textbf{Weight Modulation}}
The directest way of modulating network is to generate weights by hyper-network \cite{ha2016hypernetworks} for the  backbone network based on hybrid context. However, the backbone network usually has thousands of parameters, making it difficult to generate such high-dimension output. Even we can reduce the parameters dimension, it is still unstable for training if all the parameters are generated via hyper-network. Thus, we choose to only generate weights and bias by hyper-network $f_{\theta_h}$ for the final linear layer. In fact,  Vartak et al.\cite{vartak2017meta} also try to generate weights for a linear model. But they only conduct modulation on a single linear model rather than the last layer of MLP, which limits the representation ability of the modulated network. The detailed equation is shown as follows.
\begin{equation}
\begin{aligned}
    \label{equ: liner modulation}
    w_h, b_h &= f_{\theta_h}(C_h),\\
    o_{ui} &= w_h^T h_{ui} + b_h,
\end{aligned}
\end{equation}
where we denote the hybrid context as $C_h$, hyper-network's parameter as $\theta_h \in \Theta_M$, the generated modulation parameters as $w_h,b_h$, the output of final hidden layer as $h_{ui}$ and the final output as $o_{ui}$. \par
The strength for weight modulation is its simplicity, which makes it quite easy to implement and suitable for some simple tasks. When tasks are quite different, this modulation may not have enough representation ability and capacity for rapid adaptation. We offer the second way, layer modularion, to solve the problem.
\subsubsection{\textbf{Layer Modulation}}
As mentioned above, it is hard to directly generate weights for all hidden layer's parameters because of its high dimensionality. Therefore, we adopt layer modulation rather than layer weights' modulation, where we modulate layers ouput by hyper-network. It can effectively lower the dimensionality of parameter space from $n^2$ to $n$, where $n$ refers to the amount of nodes in one layer. Here we show two ways of layer modulation. The first one is that  hyper-network generates weights activated by Sigmoid function  and the modulated output of each layer is the dot-product of the original  output and the generated weights.  The second way follows \cite{perez2018film}(FiLM) by utilizing feature-wise Linear Modulation on backbone network. The hyper-network generates weights and bias for linear modulation on layers' output. For layer $i$'s output $l_i$, the sigmoid-dot modulation and FiLM modulation can be written by Equations (\ref{equ:sigmoid}) and (\ref{equ:film}) respectively.
\begin{equation}
\begin{aligned}
    \label{equ:sigmoid}
    w_h &= \text{Sigmoid}(f_{\theta_h}(C_h)),\\
    o_i &= w_h \odot l_i,
\end{aligned}
\end{equation}
\begin{equation}
\begin{aligned}
     \label{equ:film}
     w_h, b_h &= f_{\theta_h}(C_h),\\
    o_i &= w_h \odot l_i + b_h,
\end{aligned}
\end{equation}
where we denote the hyper-network's parameter as $\theta_h \in \Theta_M$, the generated modulation parameters as $w_h,b_h$ and the modulated output of layer $i$ as $o_i$. \par
Layer modulation increases its representation ability for task adaptation compared with the weight modulation. The weakness is that the modulation weights are still high dimensional and even though we can interpret it by visualizing the modulation weights' clusterings, it is still hard to interpret how the model works for specific tasks. That is why we provide the third modulation way called soft modularization.
\subsubsection{\textbf{Soft Modularization}}
In this section we introduce soft modularization method. The model modulation is conducted by controlling mixture of experts network. Some works like \cite{ma2018modeling} discuss how mixture of experts with shared bottom layer can be used for handling multi-task recommendation problems. The multi-task setting has some similarity with meta learning setting since both  need to adapt model when facing one specific task. For modulating the network in our setting, we adopt the similar way of soft modularization from \cite{yang2020multi} to generate route weights for sub-networks. A main difference is that our modulation is conditioned on continuous hybrid context which can easily generalize to completely new task while the task indicator used in \cite{yang2020multi} can only handle multi-task problem. The soft modularization consists of two parts: base network and route network. In base network, there exists $k$ layers and each layer  has several sub-networks (modules). For instance, there exists four $32 \times 32$ fully connected sub-networks at each layer, which has equivalent amount of parameters with a $64 \times 64$ fully connected network. Conditioned on the hybrid context, the route network will generate dynamic routing for base network, which consists of $k$ probability distributions. Each distribution is used to aggregate the output of sub-networ  k.\par
The detailed procedures are as follows: when a hybrid context is fed into the route network, it will be used to generate a probability distribution of module's route weights by Softmax activation function for each base network layer. For a $k$-layer base network with $m$ modules each layer, the route network with the same depth $k$ will generate $k$ $\mathbb{R}^{m \times m}$ matrices for all layers. In each matrix, the $i$-th ($i\in{1,2...m}$) column vector sums up to 1 and corresponds to the weights between $i$-th module at layer $l$ to all $m$ modules at layer $l+1$.  We denote the route vector logits of layer $l$ as $\sigma^l \in \mathbb{R}^{m\times m}$, the probability scalar from $i$-th module in layer $l$ to module j in layer $l+1$ as $p_{i,j}^l \in \mathbb{R}$, the output of $l$-th hidden layer as $h_r^l$, the meta model parameters as $\theta_h \in \Theta_M$, the modulated output of $i$-th modules at layer $l$ as $ob^l_i$.  Equations (\ref{equ:route}) and (\ref{equ:base}) depict how the route function and aggregated base network work in layer $l$.
\begin{equation}
 \begin{aligned}
    \label{equ:route}
    \sigma^l &= \text{MLP}_{\theta_h}(h_r^{l-1}),~~~ h_{r}^{0} = C_h,\\
    p_{i,j}^l &= \frac{\text{exp}(\sigma^l_{i,j})}{\sum_{j=0}^{m-1}\text{exp}(\sigma^l_{i,j})},
\end{aligned}   
\end{equation}

\begin{equation}
    \label{equ:base}
    ob^l_{i} = \sum_{j=0}^t p_{i,j}^{l-1}\text{MLP}_{\Phi}(ob^{l-1}_{j}),
\end{equation}\par
Soft modularization can somehow be treated as shared layer modulation, since the route weight is shared among all nodes in one module. The representation ability of soft modularization will be discounted compared with layer modulation. However, the route weights generated by route network are now low-dimensional and we can easily get access to the weight distribution to know how the sub-modules are activated for a specific task, which strongly increases the interpretability.
\begin{table}
  \caption{Time and Space complexity for gradient based and context-based Meta Learning algorithms}
  \label{tab:complexity}
  \begin{tabular}{cccl}
    \toprule
    Alogrithm&Time&Space&Process\\
    \midrule
    MAML-k & $O(mk(f_t+b_t))$& $O(mk(f_s+b_s))$&Infer\\
    MAML-k& $O(5km b_t)$&$O(2km b_s)$&Train\\
    MAML-$\delta$&$O(5mb_t \kappa \log \left(\frac{D}{\delta}\right))$& $O(2m b_s\kappa \log \left(\frac{D}{\delta}\right))$&Train\\
    CMML &$O(m f_t)$&$O(m f_s)$&Infer\\
    CMML &$O(m b_t)$&$O(m b_s)$&Train\\
  \bottomrule
\end{tabular}
\end{table}
\subsection{Computational Efficiency Analysis}
Though  the gradient based meta learning approach brings a  notably performance improvement over previous traditional recommendation algorithms, it imposes considerable computational and memory burdens  due to the the inner-loop gradient operator.
In order to show the strengths of CMML framework with respect to the computational efficiency, we briefly analyze the time and space complexity for MAML-like algorithms and CMML in training and inference phase. Note that the analysis here only considers the computational complexity for basic MAML based algorithms like \cite{lee2019melu}. \cite{du2019sequential} and \cite{dong2020mamo} will have larger computational complexity. In the following analysis, we denote $f_t, f_s, b_t, b_s$ as the time and space complexity for forward process and backward process respectively. Assume we have $m$ data points to be fed into the neural network and we take k gradient steps in the inner-loop.\par
In the inference phase, the MAML-like algorithms need to perform feed-forward operation and backward operation for m data points each step, so the final time and space complexity for MAML-like algorithms is $O(mk(f_t+b_t))$ and $O(mk(f_s+b_s))$. For CMML, it only needs to take feed-forward operation, resulting in the final time and space complexity: $O(m f_t)$, $O(m f_s)$. Note that in most cases, $b_t > f_t$ and $b_s > f_s$ holds. So the time and space complexity of MAML-like algorithms is 2k times larger than that of CMML in the inference phase.\par
In the training phase, the efficiency strength of CMML can be more clear since the gradient based method needs to differentiate through the whole gradient paths in the inner-loop, which requires the computation of the hessian-vector product. In this part, we neglect the complexity analysis for forward process. \cite{rajeswaran2019meta} and \cite{shaban2019truncated} show that in the reverse mode of gradient calculation, the time and space complexity of hessian-vector product are typically no more than a constant over that of first-order gradient calculation. Usually, the constant is 5 and 2 for time and space complexity respectively. Thus, for MAML-like algorithms, the time and space complexity is $O(5km b_t)$ and $O(2km b_s)$, while for CMML,  the time and space complexity is $O(m b_t)$ and $O(m b_s)$. \par
The analysis is conducted given fixed gradient steps k. Combining our analysis and the Prop 3.1 of \cite{shaban2019truncated}, we can also show that when we are trying to get a $\delta-\text{accurate}$ optimal solution in the inner-loop (which means the distance between optimized parameters and optimal parameters of the inner-loop problem is bounded by $\delta$), the time and space complexity is $O(5mb_t \kappa \log \left(\frac{D}{\delta}\right))$ and $O(2m b_s\kappa \log \left(\frac{D}{\delta}\right))$, where $\kappa$ is the condition number for inner-loop optimization and $D$ is the diameter of parameters' search space. We summarize the analysis results in Table \ref{tab:complexity}.

\section{EXPERIMENT}
\label{experiment}
In this section, we detail the experimental settings and results to validate the effectiveness of CMML. This section includes the experimental settings like dataset configuration, evaluation metrics and baseline algorithms. The performance comparison section shows that CMML can achieve both outstanding performance and high computational efficiency. Visualization of the learned embedding and route is shown for demonstrating the interpretability of CMML. And finally, the ablation study is conducted to analyze how specific structure influences the performance of CMML.
\subsection{Experimental Settings}
\subsubsection{Dataset} For scenario-specific recommendation setting, we evaluate our algorithm on two public datasets from MovieLens-20M \cite{harper2015MovieLens} and Taobao \cite{du2019sequential}.   MovieLens-20M is a large movie-rating dataset from movie recommendation service MovieLens, and it consists of 138,493 users, 27,279 movies, and 20,000,263 rating records. Following similar CTR problem formulation of \cite{du2019sequential}, we transform the MovieLens-20M dataset from explicit rating into implicit feedback. All movies rated by the user  will be regarded as positive samples with the rest as negative samples. 306 different genres of movie are treated as scenarios for scenario-specific setting. The second Taobao dataset\footnote{Downloadable from https://tianchi.aliyun.com/dataset/dataDetail?dataId=9716.} is from the cloud theme click log on Taobao recommender system. This dataset consists of users' purchase history in 355 different themes such as \textit{"things to prepare when a baby is coming"}, which are treated as different scenarios. There are 700k users, 1400k items, and 5717k purchase history in Taobao. We also build a synthetic dataset called hybrid Movie-Taobao which concatenates scenarios in both datasets to test our algorithms on multi-domain cold-start problems. The meta-training and meta-testing dataset of all three datasets are split based on different scenarios.\par
For user-specific recommendation setting, we use the MovieLens-1M\cite{harper2015MovieLens} dataset following \cite{lee2019melu}. MovieLens-1M is an older and smaller version of MovieLens dataset compared with MovieLens-20M. It consists of 6,040 users, 3,706 movies, and 1,000k rating records. Different from the CTR problem in scenario-specific setting, we need to solve the cold-start regression problem for user-item score prediction. The meta-training and meta-testing dataset of it are split based on different users, denoted as $W-W$ and $C-W$ (which is short for two settings of training on warm users - testing on warm users and training on cold users - testing on warm users respectively). Note that the split method is different from that in \cite{lee2019melu} because cold users are defined as earlier users than warm users and meta learning problem is motivated to the goal of  generalizing towards unseen or new tasks (users).\par
Here we also detail the data preprocessing procedures. In scenario-specific setting, we only select scenarios with less than $1000$ but more than $100$ items in each dataset to make sure it can guarantee the cold-start property. There exists 232 scenarios for meta-training set with the rest as meta-testing set. In the training process, we will randomly sample $64$ user-item positive pairs as support-set and $128$ user-item pairs as query-set from each scenario. Note that since the amount of negative samples is much larger than that of positive samples, we will also sample the same amount of negative pairs in support and query-set. For user-specific setting, we trim the dataset and only choose $50$ rating records for each user to match the cold-start setting. There exists 4832 users in meta-trainng set with the rest as meta-testing set. For each user, we randomly sample $10$ items for query-set with the rest as support-set.\par
\subsubsection{Evaluation Metrics}
For scenario-specific setting, since it is modeled as a CTR prediction problem in \cite{du2019sequential}, we adopt the same metric and utilize top-N recall for performance evaluation. And for user-specific setting, which is modeled as a score regression problem in  \cite{lee2019melu}, we utilize  both mean absolute error (MAE) and normalized discounted cumulative gain (NDCG\cite{jarvelin2017ir}) for evaluation as used in  \cite{lee2019melu}, and the higher the better in NDCG,
 the lower the better in MAE. The experimental results also reveal that our method generally works for different loss functions.
\subsubsection{Baseline algorithms} We select many state-of-the-art baselines of deep learning or meta learning based recommendation algorithms for comparison. For scenario-specific setting,
\begin{itemize}
    \item \textbf{Deep Cross Network(DCN)}: Deep cross network \cite{wang2017deep} is one of the most classical item recommendation algorithms and it has been widely applied in industrial recommendation system. It proposes cross layer which successfully combines benefits of FM and DNN for item recommendation. We train our DCN model on all scenarios in the meta-training set.
    \item \textbf{Deep cross network with fine-tuning(DCN-F)}. We also provide a baseline which finetunes the pretrained DCN model on specific scenario based on data in the support-set.
    \item \textbf{ItemPop}: ItemPop is a heuristic algorithms used in \cite{rendlebpr} which ranks items according to its popularity, measured by the amount of corresponding interactions in the support-set.
    \item \textbf{CoNet}: CoNet is a cross-domain recommendation algorithms for tackling cold-start problem. It utilizes cross connection to enable dual knowledge from source domain to target domain. We reimplement the algorithm on MovieLens-20M by regarding movies without labeled genre as the source domain. For Taobao datast and hybrid dataset, the authors did not offer the source domain data used in \cite{du2019sequential}. Thus here we can only show the reported result in \cite{du2019sequential} in Taobao dataset.
    \item \textbf{$s^{2}$ Meta}\cite{du2019sequential}: The  state-of-the-art meta learning approach for scenario-specific cold-start problem. It proposes update and stop controllers for better inner-loop optimization.
\end{itemize}
For user-specific setting, 
\begin{itemize}
    \item \textbf{DCN} and \textbf{DCN-F}: Same with previous baselines and we modify the model into a regression model.
    \item \textbf{MetaCS-DNN}\cite{bharadhwaj2019meta}: A meta-learning baseline that tries to tackle the cold-start user recommendation based on MAML-like algorithms. For a specific user, the whole model will take a few gradient steps for adaptation.
    \item \textbf{MELU} \cite{lee2019melu}: MELU basically shares similar idea with MetaCS-DNN by utilizing MAML-like algorithm. The only difference is the inner-loop optimization will only happen in fully connected layers rather than the whole network.
\end{itemize}

\subsubsection{Features and Hyper-parameters}
For MovieLens-1M dataset, we select the same features as used in \cite{lee2019melu}, and for MovieLens-20M dataset and Taobao dataset, we seclect the same features as adopted in \cite{du2019sequential}. The network architectures involved are designed to be comparable with the baselines for fairness. The base network is constructed as a ReLU activated MLP with hidden units  $(64,64,64)$, the context encoder includes a GRU and a ReLU activated MLP  with hidden units $(128,128)$. For CMML-FiLM, we also utilize a ReLU activated MLP with hidden units $(64,64,64)$ for layer modularization. For CMML-Soft-M, it contains $k=3$ module layers, $m=4$ modules per layer and each module outputs a $d = 32$ representation. The user embeddings and item embeddings are pre-generated \cite{du2019sequential} by collaborative filtering \cite{koren2009matrix}. We optimize our model with Adam \cite{kingma2014adam} optimizer with learning rate 1e-4. These hyper-parameters are configured by grid search.

\textbf{Clarification: During our preparation for code, we find out in the scenario-based experiments, the random seeds have some influences on our final experimental results. So we reconduct experiments in the scenario-based setting for 5 seeds and report the results in the appendix.}
\subsection{Performance Comparison}
\begin{table*}[t]
\caption{Top-N recall in 3 datasets}
\label{tab:scenario}
\centering
\begin{tabular}{l|ccc|ccc|ccc}
\hline  & \multicolumn{3}{|c|} { MovieLens-20M } & \multicolumn{3}{|c} { Taobao } & \multicolumn{3}{|c} { Hybrid Movie-Taobao}\\
\cline { 2 - 10 } Method & Recall@10 & Recall@20 & Recall@50 & Recall@20 & Recall@50 & Recall@100 & Recall@20 & Recall@50 & Recall@100\\
\hline DCN & 31.50 & 52.27 & 85.73 & 23.52 & 39.21 & 56.27 & 35.12 & 57.13 & 72.05\\
DCN-F & 41.08 & 60.59 & 86.83 & 20.69 & 35.56 & 52.26 & 37.30 & 55.85 & 69.75\\
ItemPop & 39.65 & 54.33 & 78.12 & 21.94 & 36.11 & 38.19 & 35.00 & 52.76 & 61.02\\
CoNet & 46.27 & 63.11 & 86.98 & 20.27 & 31.48 & 44.53 & - & - & -\\
s $^{2}$ Meta &  $\mathbf{4 7 . 52}$ & $\mathbf{6 6 . 09}$ & $\mathbf{8 9 . 08}$ & 25.52 & 42.03 & 58.12 & 39.36 & 56.84 & 70.19\\
\hline
CMML-FiLM &  47.36 & 65.32 & 88.69 & $\mathbf{2 5 . 77}$ & $\mathbf{4 2 . 30}$ & $\mathbf{5 8 . 77}$ & $\mathbf{40.74}$ & $\mathbf{59.92}$ & $\mathbf{73.85}$\\
CMML-Soft-M &  47.32 & 65.62 & 88.81 & $\mathbf{2 5 . 56}$ & $\mathbf{4 2 . 19}$ & $\mathbf{5 8 . 66}$ &$\mathbf{40.47}$ & $\mathbf{59.94}$ & $\mathbf{73.76}$\\
\hline
\end{tabular}
\end{table*}
\begin{table}[t]
\centering
\caption{Performance in MovieLens-1M}
\label{tab:user}
\begin{tabular}{l|cc|cc}
\hline  & \multicolumn{2}{|c|} { W-W } & \multicolumn{2}{|c} { C-W }\\
\cline { 2 - 5 } Method & MAE  & NDCG@3 & MAE  & NDCG@3\\
\hline DCN & 0.869 & 0.693 & 0.907 & 0.665 \\
DCN-F & 0.831 & 0.683 & 0.847 & 0.679 \\
MetaDNN & $\mathbf{0.726}$ & $\mathbf{0.772}$ & $\mathbf{0.733}$ & $\mathbf{0.765}$ \\
MELU & $\mathbf{0.725}$ & $\mathbf{0.771}$ & $\mathbf{0.734}$ & $\mathbf{0.763}$\\
\hline
CMML-FiLM & $\mathbf{0.727}$ & $\mathbf{0.771}$ & 0.745 & $\mathbf{0.760}$\\
\hline
\end{tabular}
\end{table}
In this part we summarize the performance comparison for CMML (based on FiLM layer modulation and soft modularization) and other baseline algorithms in scenario/user-specific setting, respectively.\par
For scenario-specific setting,  Table \ref{tab:scenario} shows that CMML generally achieves comparable or better results among all three datasets in terms of top-N recall. 
For Taobao dataset, we notice that DCN-F and other scenario-specific algorithms except $s^2$ Meta perform worse than DCN. The counterintuitive phenomenon reveals that scenario-specific support-set in Taobao may contain noisy and irrelevant information, which might be detrimental since adaptation may easily cause over-fitting on limited noisy support-set samples. Under this condition, CMML still achieves improvement comparable to the best baseline $s^2$ Meta, which validataes the effectiveness of CMML. The results also reveal that CMML can work for different scales of datasets. Hybrid MovieLens-Taobao evaluates the algorithms' performance on datasets composed of multi-modal data. CMML achieves much improvement over the best baseline $s^2$ Meta. The reason might be that gradient based algorithms can only lean one group of initial parameters, which can work pretty well dealing with single-domain dataset, yet not suitable for multi-domain dataset. CMML works better here due to its flexible representation and modulation ability for multi-domain datasets.\par

For user-specific setting,  we basically observe similar results. The fine-tuning of DCN-F helps for identifying users preferences and increasing the performance. Three meta learning based algorithms achieve comparable results and are better than DCN baselines.\par

We remark that CMML does not achieve remarkably better results on single modal datasets compared with gradient based method. That is because the gradient descent is a strong baseline for adaptation and can guarantee performance improvement even without any meta-training. We leave further investigation of combining CMML with gradient based methods for future work.

\begin{figure}[t]
    \centering
    \includegraphics[width=0.48\linewidth]{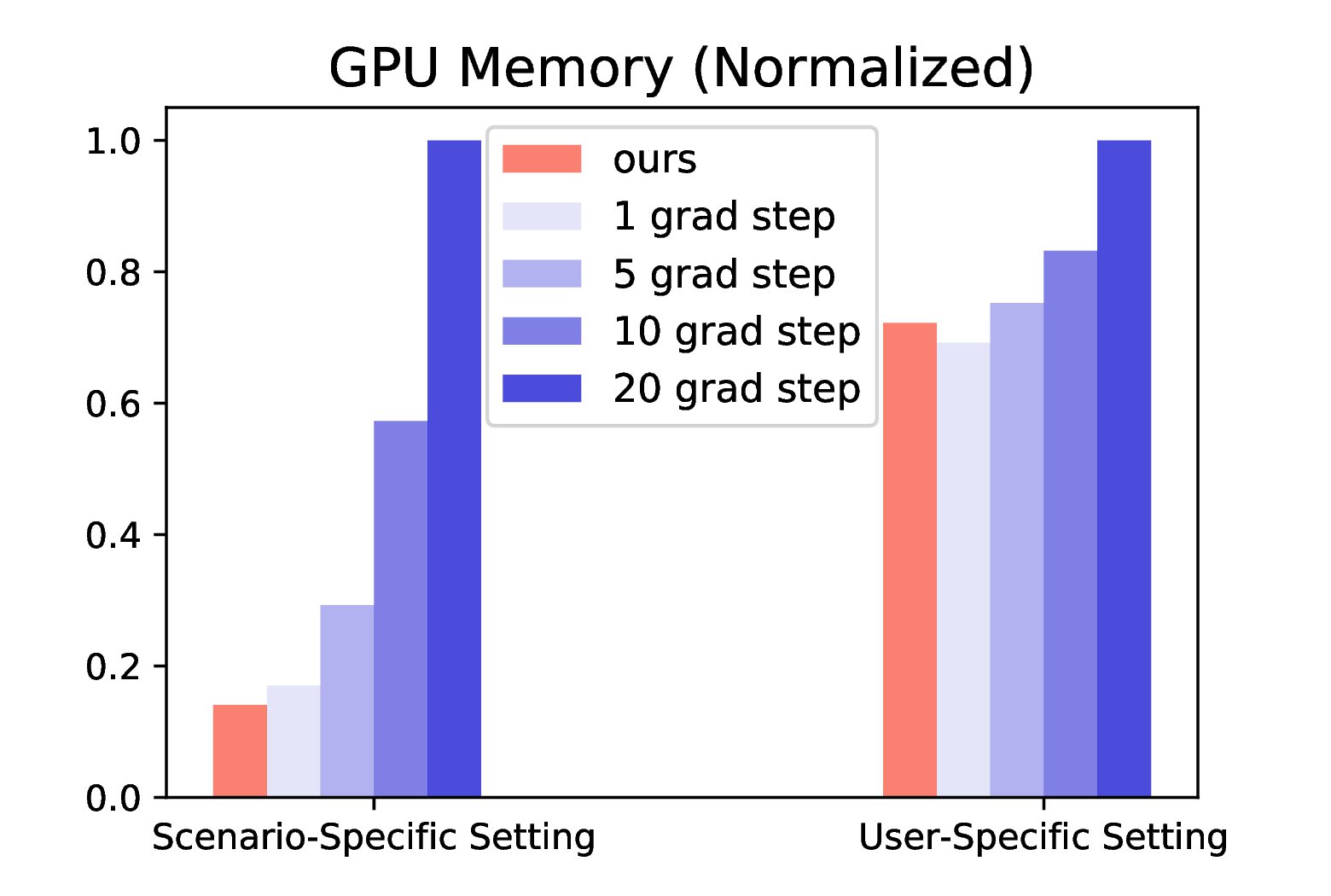}
     \includegraphics[width=0.48\linewidth]{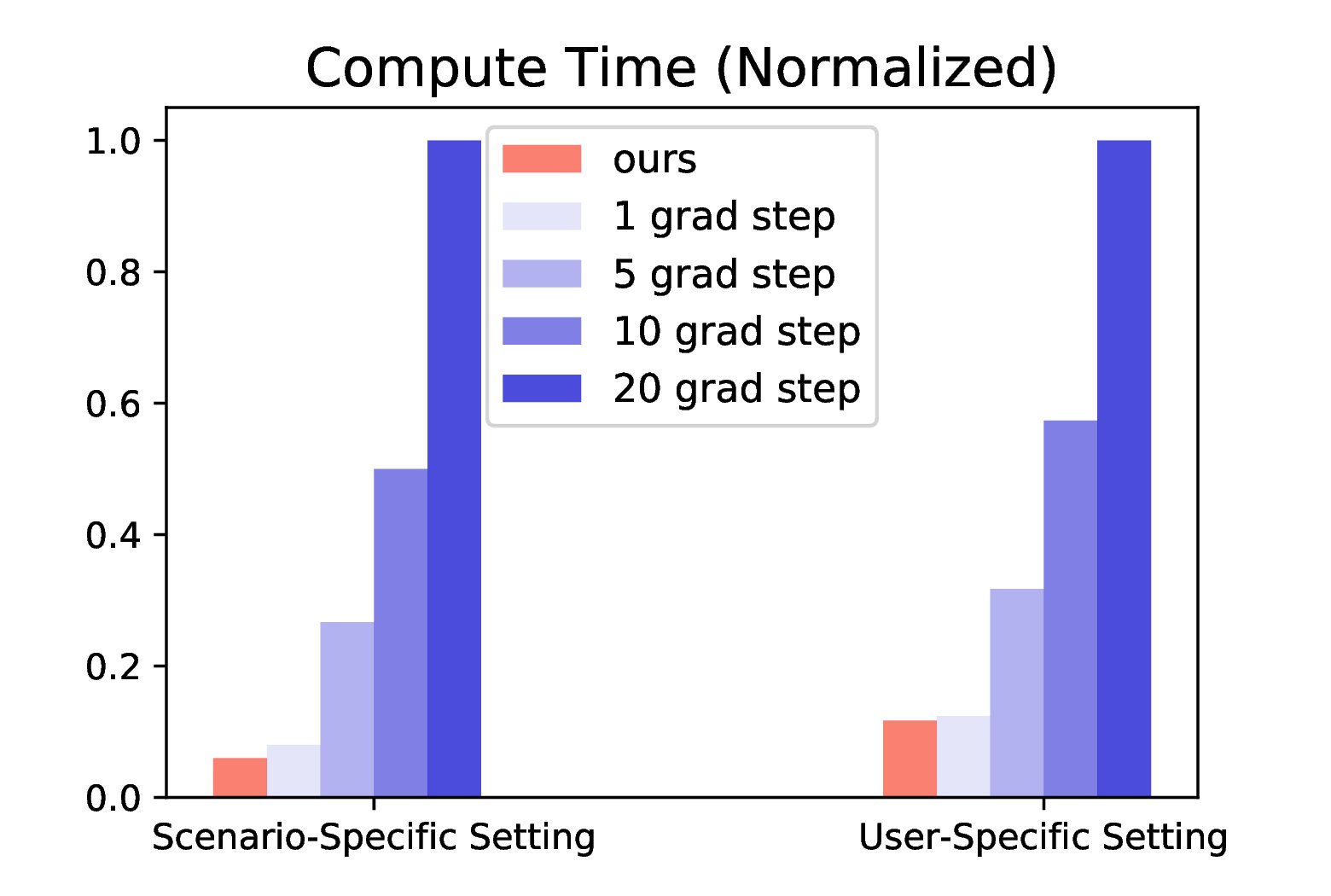}
    \caption{Memory and time}
    \label{fig:time}
\end{figure}
\begin{figure}[t]
 \centering
 \includegraphics[width=0.48\linewidth]{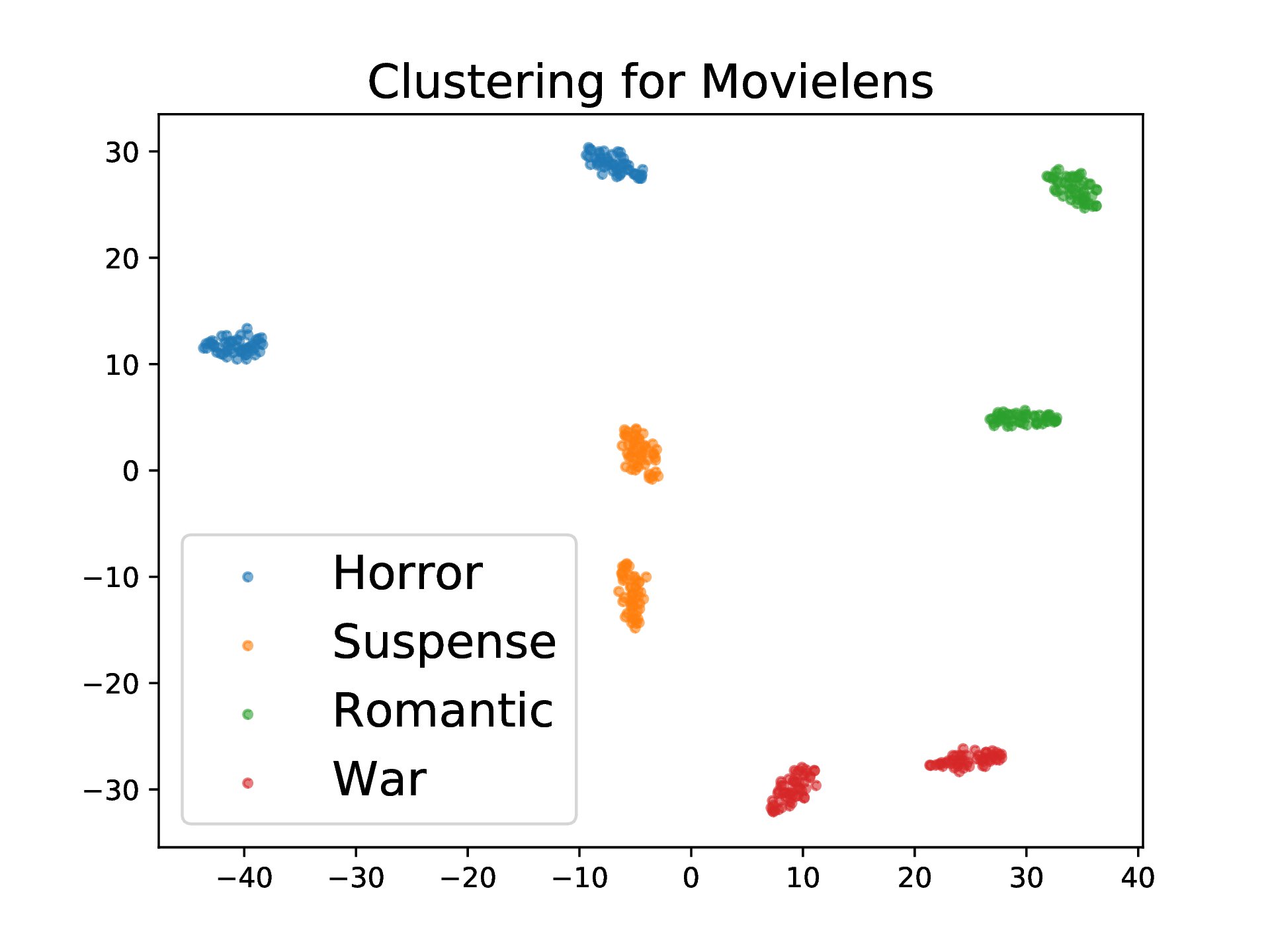}
 \includegraphics[width=0.48\linewidth]{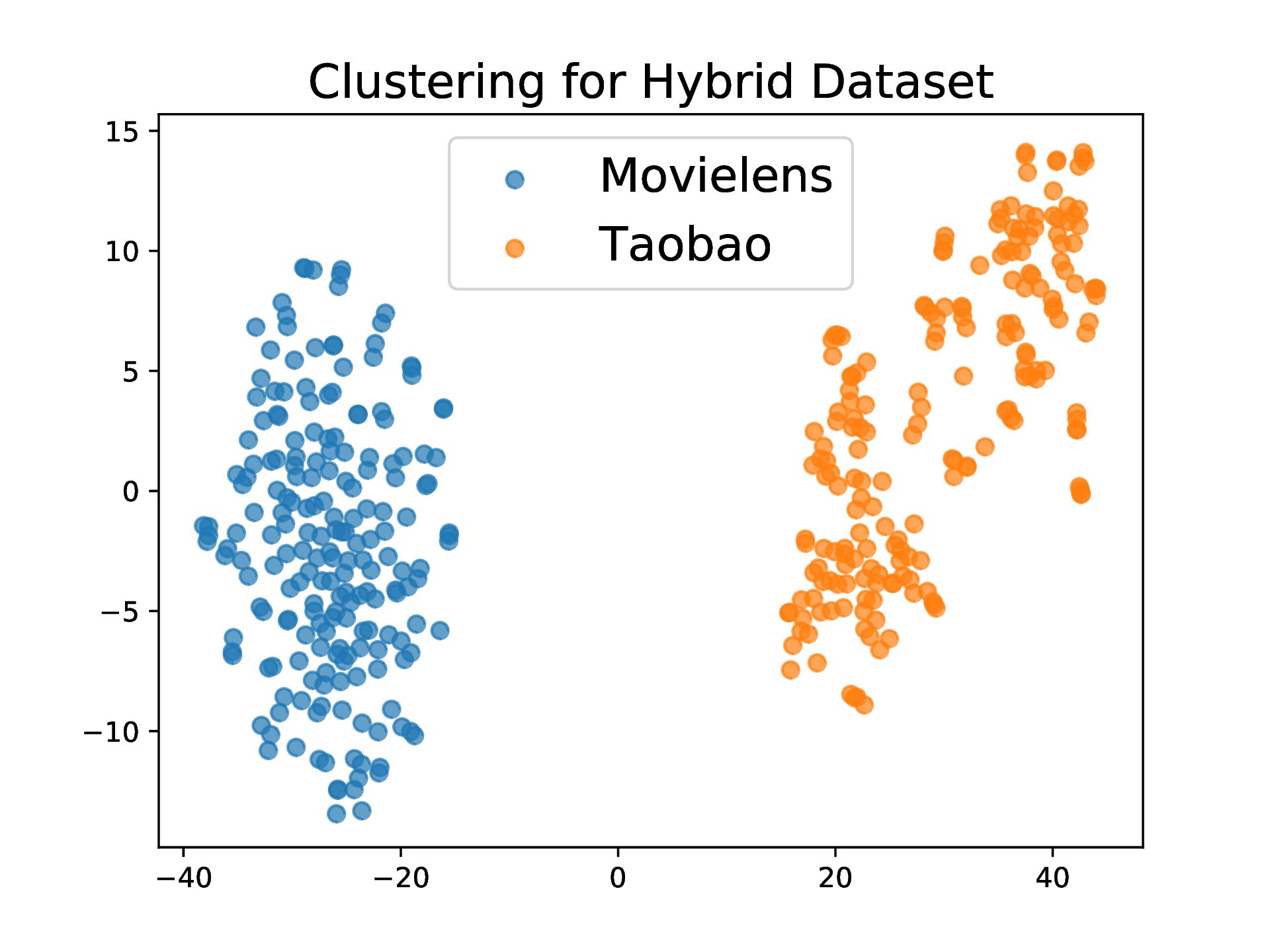}
\caption{Visualization of task-level context}
\label{fig:vi}
\end{figure}

We also provide the comparison results of  time and memory efficiency of $s^2{\text{Meta}}$, MELU and CMML in Figure \ref{fig:time}. We use sequential encoder, dot hybrid context generator and FiLM based modulation network for CMML, and the inner gradient steps of $S^2_\text{Meta}$ and MELU are both set with $1, 5, 10, 20$. It can be shown that the time and memory consumption increases with more gradient steps. For scenario specific setting, our algorithm achieves remarkable computational efficiency compared with $S^2\text{Meta}$. It consumes less memory and time compared with even one-step $S^2\text{Meta}$ while it usually conducts 20-step gradient descent in the implementation.  For user-specific setting, our algorithm achieves comparable time and memory demands with one-step MELU, while MELU needs to take 5-step gradient descent in implementation. Note that here CMML has more model's parameters compared with $S^2\text{Meta}$ and MELU because of all three additional modules, but it is still quite computationally efficient.
\begin{figure}[t]
    \centering
    \includegraphics[width=0.6\linewidth]{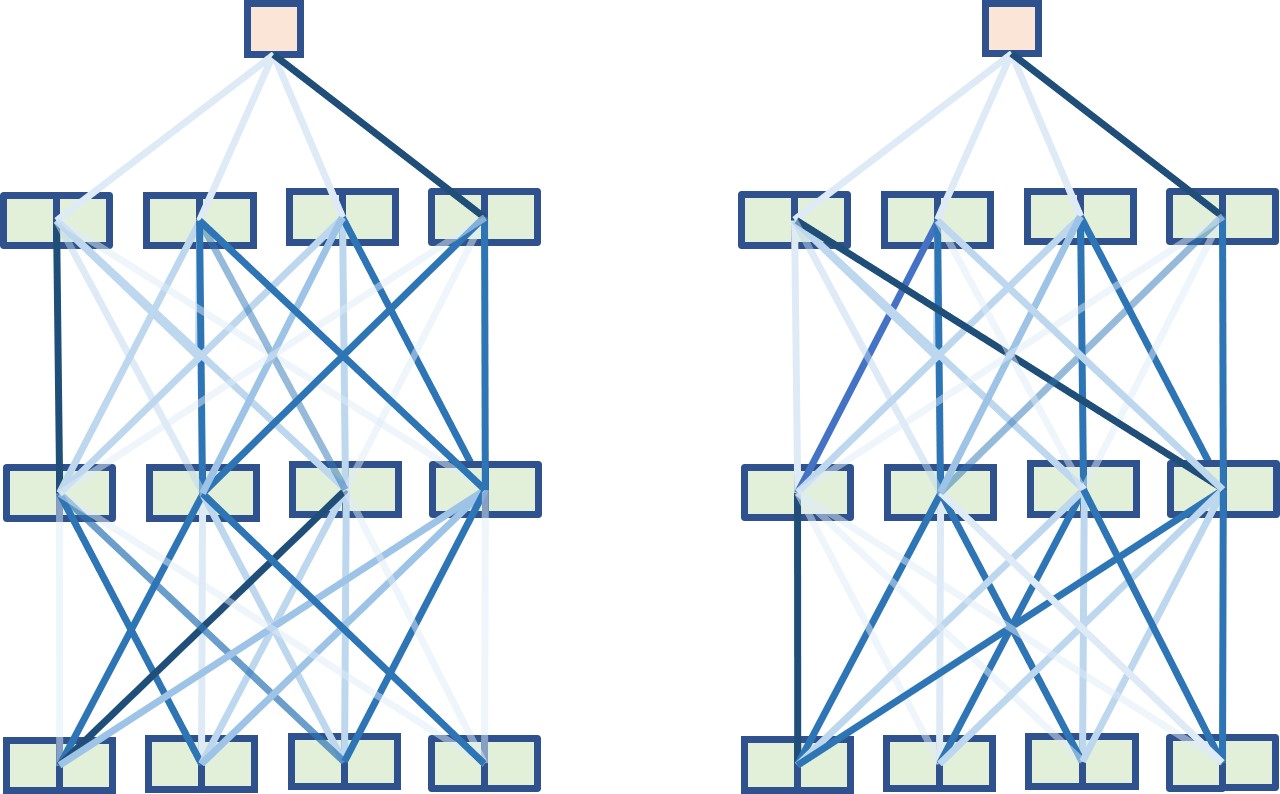}
    \caption{Activated routes for two different tasks}
    \label{fig:wv}
\end{figure}
\subsection{Embedding and Route Visualization}
In this section, we provide the visualization of learned context embedding and the activated route of soft modularization to show the interpretability of our proposed CMML in  Figure \ref{fig:vi} and  \ref{fig:wv} respectively. In Figure \ref{fig:vi}, we show the visualization for task-level context on MovieLens-20M dataset and hybrid dataset. For the clusterings on MovieLens, we sample the support-set for $8$ different genres(tasks) and the corresponding context embedding can be successfully clustered. It demonstrates that the context encoder is capable of distinguishing movies from different genres. In addition, we investigate the clustering of semantic groups by classifying these $8$ genres into $4$ broader types: horror, suspense, romance and war. We can find that the tasks with semantically similar genres (shown with the same color) are closer, revealing that context encoder can learn some implicit semantic information from the dataset. The second figure shows the context embedding clustering within the hybrid dataset. All task context embeddings are divided into two clusters which exactly correspond to two datasets - MovieLens20M and Taobao. In Figure \ref{fig:wv}, we display the activated route visualization of two different tasks for soft modularization, where darker color represents higher probability. Due to the low-dimension property of soft modulairzation, we can easily know what modules are activated for a specific task.\par
The task-level context and route visualization can successfully validate the appealing property of better interpretability of CMML.
\subsection{Ablation Study}
In this section, we conduct ablation study to illustrate the effectiveness of the context encoder, hybrid context generator, and context modulation network.\par
In Table \ref{tab:ch}, we show the ablation results for different context encoder and hybrid context generator on Moviesen20M dataset. All results here are based on final weight modulation. Firstly, in order to validate the importance of useful interaction history, we conduct one setting called 'Non-negative-PE', which means we remove the negative user-item pairs in the support-set and utilize pooling aggregated encoder. The comparison between 'Non-negative-PE' and 'PE' shows without useful interaction history, it will be hard to generate scenario-specific context and further modulate the network. The comparison between pooling aggregated encoder(PE) and sequential aggregated encoder(SE) shows context extraction requires high representation ability for context encoder. The comparison between SE and SE plus hybrid MLP generator shows the importance of building hybrid context for specific instance example. And between hybrid MLP generator and hybrid dot generator, the latter one is not only free of additional parameters, but also achieves better performance. Low-order interaction is enough to build the relationship between task-level context and instance's features. Thus, we set sequential aggregated encoder plus hybrid dot generator as the default setting for CMML.\par
\begin{table}[t]
    \centering
    \caption{Context encoder and hybrid context generator}
    \label{tab:ch}
    \begin{tabular}{cccc}
    \hline Method & Recall@10 & Recall@20 & Recall@50 \\
    \hline Non-negative & 31.67 & 52.69 & 86.20 \\
     PE& 40.05 & 61.24 & 87.78 \\
     SE& 44.30 & 63.82 & 88.10 \\
     SE-MLP& 46.37 & 65.18 & 88.61 \\
     SE-dot& $\mathbf{47.29}$ & $\mathbf{65.50}$ & $\mathbf{88.62}$ \\
    \hline
    \end{tabular}
\end{table}
\begin{table}[t]
    \centering
    \caption{Results for different Modulation Network}
    \label{tab:module}
    \begin{tabular}{l|cc|cc}
    \hline  & \multicolumn{2}{|c|} { MovieLens-20M } & \multicolumn{2}{|c} { Taobao }\\
    \cline { 2 - 5 } Method & Recall@10 & Recall@20 & Recall@20 & Recall@50 \\
\hline Linear & 47.29 & 65.50 & 24.63 & 41.30 \\
S-LM & 47.21 & 65.39 & 24.68 & 41.39 \\
F-LM &  $\mathbf{47.36}$ & 65.32 & $\mathbf{25.77}$ & $\mathbf{42.30}$ \\
Soft-M & $\mathbf{47.32}$ & $\mathbf{65.62}$ & $\mathbf{25.56}$ & $\mathbf{42.19}$ \\
    \hline
    \end{tabular}
\end{table}
Table \ref{tab:module} shows the ablation study on different approaches for modulating the model on MovieLens-20M dataset and Taobao dataset. The method linear correpsonds linear weight modulation. S-LM and F-LM denote Sigmoid based and FiLM based layer modulation respectively. And Soft-M represents the soft modulation approach. In MovieLens-20M dataset, all modulation ways except S-LM achieves comparable results. In Taobao dataset, F-LM and Soft-M achieve better results compared with rest two approaches. The complexity of dataset will influence the need for modulation network's capacity. For relatively simple Movielen dataset, the ability of linear weight modulation is enough to handle the model adaptation, while in Taobao dataset, stronger network capacity for the model adaptation is required,  and that' why F-LM and Soft-M achieve better results. 
\section{RELATED WORK}
\label{related work}
\subsection{Contextual Recommendation}    
There are many works \cite{adomavicius2011context, van2013deep, adomavicius2005incorporating,sugiyama2004adaptive,white2010predicting,bennett2012modeling} showing that by leveraging extra explicit contextual information like time, location, or users' profile, the recommendation algorithms can improve performance. Compared with traditional algorithms that map user-item pairs into some scores, context aware recommendation framework will take additional context information as input. The difference between our framework and context-aware recommendation is that we do not have access to any explicit contextual information. We believe the historical interaction information itself is already a great source of contextual information which can reveal characteristic of the recommendation entities. Our solution is also somewhat conceptually related with the domain-adaptation problem, where they focus on domain transfer from source domain to target domain, and invariant contextual invariant information is extracted  to achieve domain transfer \cite{DBLP:conf/sigir/KrishnanDBYS20}, \cite{DDTCDR}. CMML focuses more on the few shot cold-start recommendation problem. \par
\subsection{Meta Learning for Recommendation}
Meta Learning, also known as learning to learn, is a new rising machine learning paradigm for learning meta knowledge. Specifically, the meta learning framework aims at learning useful inductive bias, which is helpful when the available data is limited. There are many applications of meta learning, like \cite{vinyals2016matching} for one-shot classification in computer vision, and \cite{rakelly2019efficient} for few-shot meta reinforcement learning problem. One of the most classical algorithms is MAML \cite{finn2017model}, which aims at learning model's initial parameters which are capable of adapting to new tasks with only a few gradient steps.\par
There exists work \cite{vartak2017meta} that incorporates the meta learning into recommendation problem. Many recent works introduce the framework of MAML into the cold-start recommendation problems. MELU \cite{lee2019melu} introduces the MAML framework into user-specific cold-start recommendation problems, in which it transforms the cold start recommendation problem for new coming users/items as new coming tasks in the setting of MAML. $S^2$ Meta \cite{du2019sequential} tries to tackle the cold-start problem of scenario setting by learning proper initial MLP parameters, better gradient controller, and an RL stop-controller for better inner-loop adaptation. MAMO \cite{dong2020mamo} introduces task-specific memories and
feature-specific memories to get rid of problems brought by global parameter sharing in MELU. \cite{wei2020fast} equips MAML with dynamic subgraph sampling, which can solve the problem for dynamic arrival of new users. \cite{zhang2020retrain} models the training process of recommender system as a meta learning problem and propose a framework for meta learning model retraining mechanism. \cite{zheng2020cold} generates sequential recommendation results by similarity metric between query-set and support-set.
\section{CONCLUSION}
\label{conclusion}
In this paper, we propose a contextual modulation meta learning framework for solving cold-start recommendation problems with several algorithmic alternatives. We formulate the cold-start recommendation problem as a meta learning problem. By context encoder, hybrid context generator and modulation network, our framework can easily adapt to new tasks even with limited interaction examples. Extensive experiments on real-world datasets successfully validate that effectiveness of CMML with much higher computational efficiency and interpretability. Also, the whole framework is completely compatible with the current practical industrial framework for broader application.
\bibliographystyle{ACM-Reference-Format}
\bibliography{sample-base}


\section{Appendix}
\subsection{Multi-seed results}
For reproducibility and more reliable results, we conduct experiment on 5 seeds for the scenario-based setting and report the average and the confidence interval (calculated by $\text{std} * 1.96 /\sqrt{5}$).
\begin{table}[H]
    \centering
    \caption{MovieLens-20M with 5 seeds}
    \label{tab:ch}
    \begin{tabular}{cccc}
    \hline Method & Recall@10 & Recall@20 & Recall@50 \\
    \hline
    DCN&$31.25\pm0.12$&$52.92\pm0.18$&$86.19\pm0.08$\\
    DCN-F&$39.18\pm0.19$&$59.39\pm0.24$&$86.32\pm0.08$\\
    ItemPop &$39.65\pm0.00$&$54.33\pm0.00$&$78.12\pm0.00$\\
    CoNet & $46.37\pm0.08$ & $63.93\pm0.12$ & $87.29\pm0.05$\\
    s $^{2}$ Meta &  $\mathbf{4 7 . 56\pm0.02}$ & $\mathbf{6 6 . 09\pm0.04}$ & $\mathbf{8 9 . 08\pm0.02}$ \\
    \hline
     CMML-FiLM &  $47.06\pm 0.07$ & $65.14\pm 0.04$ & $88.48\pm0.02$ \\
     CMML-Soft-M &  $\mathbf{47.31\pm 0.05}$ & $\mathbf{65.65\pm 0.04}$ & $\mathbf{88.75\pm0.02}$ \\
     CMML-Linear &$\mathbf{47.20\pm 0.07}$ & $65.52\pm 0.05$ & $\mathbf{88.66\pm0.02}$\\
     CMML-Sigmoid &$47.16\pm 0.05$ & $65.49\pm 0.08$ & $88.55\pm0.03$\\
    \hline
    \end{tabular}
\end{table}
\begin{table}[H]
    \centering
    \caption{Taobao with 5 seeds}
    \label{tab:ch}
    \begin{tabular}{cccc}
    \hline Method & Recall@20 & Recall@50 & Recall@100 \\
    \hline 
    DCN&$22.58\pm0.14$&$38.88\pm0.13$&$55.05\pm0.20$\\
    DCN-F&$22.57\pm0.23$&$37.67\pm0.14$&$53.77\pm0.24$\\
    ItemPop &$21.94\pm0.00$&$36.11\pm0.00$&$38.19\pm0.00$\\
    CoNet & 20.27& 31.48& 44.53\\
    s $^{2}$ Meta &  $\mathbf{24.99\pm0.22}$ & $40.98\pm0.16$ & $57.30\pm0.14$ \\
    \hline
     CMML-FiLM &  $\mathbf{25.24\pm 0.13}$ & $\mathbf{41.84\pm 0.14}$ & $\mathbf{58.42\pm0.17}$ \\
     CMML-Soft-M &  $\mathbf{25.26\pm 0.08}$ & $\mathbf{42.07\pm 0.04}$ & $\mathbf{58.66\pm0.09}$ \\
     CMML-Linear &$24.77\pm 0.15$ & $41.49\pm 0.18$ & $58.09\pm0.14$\\
     CMML-Sigmoid &$\mathbf{25.29\pm 0.14}$ & $\mathbf{42.05\pm 0.13}$ & $\mathbf{58.38\pm0.11}$\\
    \hline
    \end{tabular}
\end{table}
\begin{table}[H]
    \centering
    \caption{MovieLens-Taobao with 5 seeds}
    \label{tab:ch}
    \begin{tabular}{cccc}
    \hline Method & Recall@20 & Recall@50 & Recall@100 \\
    \hline 
    DCN &$34.06\pm0.67$&$56.80\pm0.40$&$71.77\pm0.32$\\
    DCN-F &$36.60\pm0.50$&$55.66\pm0.27$&$69.56\pm0.36$\\
    ItemPop &$35.00\pm0.00$&$52.76\pm0.00$&$61.02\pm0.00$\\
    s $^{2}$ Meta &  $39.35\pm0.30$ & $58.01\pm0.17$ & $70.88\pm0.18$ \\
    \hline
     CMML-FiLM &  $\mathbf{40.52\pm 0.25}$ & $\mathbf{59.94\pm 0.08}$ & $\mathbf{73.77\pm0.10}$ \\
     CMML-Soft-M &  $\mathbf{40.75\pm 0.08}$ & $\mathbf{60.02\pm 0.10}$ & $\mathbf{73.89\pm0.09}$ \\
    \hline
    \end{tabular}
\end{table}
Firstly, from the new multi-seed results in Table 6, 7, 8, we can find out that one of CMML variant CMML-FiLM is sensitive to different random seeds so we can observe performance decrease on both MovieLens-20M and Taobao dataset, while CMML-Soft-M achieves really stable results and is basically consistently with the results presented in the main paper. The performance of the strongest baseline $s^{2}\text{Meta}$ is stable on MovieLens-20M but decreases a lot on Taobao. The main conclusion stays the same that we can achieve comparable results on MovieLens-20M dataset and better performance on Taobao (Especially on Recall@50 and Recall 100) and hybrid dataset (in all metrics).

In addition, we also conduct full multi-seed ablation experiments on two variants: CMML-Sigmoid and CMML-Linear. We observe similar results with the main paper on CMML-Linear. It can perform well on MovieLens-20M but not on Taobao dataset. That validates our conclusion that Taobao dataset requires stronger network capacity. The CMML-sigmoid results are different from the main paper, we find out it is more stable than CMML-FiLM variant for multi-seed experients and can achieve comparable results with CMML-Soft-M.

We believe the main reason that CMML-FiLM is sensitive to different seeds is that its modulation is the most general and powerful one without any constraints. In contrast, CMML-Sigmoid has constriants on modulation weights by Sigmoid activation function and CMML-Soft-M modulates the network with shared weight modulation. During the training, the modulation way of CMML-FiLM might be sensitive to different network initialization and training data brought by different random seeds.

\begin{table}[H]
    \centering
    \caption{Context encoder and hybrid context generator}
    \label{tab:ch}
    \begin{tabular}{cccc}
    \hline Method & Recall@10 & Recall@20 & Recall@50 \\
    \hline Non-negative & $32.83\pm0.98$ & $54.35\pm0.90$ & $86.49\pm0.16$ \\
     PE& $37.35\pm0.57$ & $58.85\pm0.51$ & $87.34\pm0.09$ \\
     SE& $41.24\pm0.66$ & $61.94\pm0.43$ & $87.95\pm0.08$ \\
     SE-MLP& $46.62\pm0.07$ & $65.25\pm0.07$ & $\mathbf{88.68\pm0.04}$ \\
     SE-dot& $\mathbf{47.20\pm0.07}$ & $\mathbf{65.52\pm0.05}$ & $\mathbf{88.66\pm0.02}$ \\
    \hline
    \end{tabular}
\end{table}
We also conduct multi-seed experiments on context encoder and hybrid context generator ablation study. Basically the conclusion stays the same with the main paper.
\subsection{Implementation Details}
For better reproducibility, in this subsection we detail how we implement our method and baseline algorithms. You can email xidong.feng.20@ucl.ac.uk if you need codebase for this paper.\par
\subsubsection{scenario-specific setting}
For scenario-specific setting, we follow \cite{du2019sequential} hyper-parameter setting for ItemPoP and CoNet baseline. For DCN baseline we use 3-layer cross network and deep network with [64,64] hidden size and ReLU activation function. We use Adam optimizer with lr=4e-7, weight decay=1e-5 and batch size=128. For DCN-F baseline we conduct hyper-parameter search over fine-tuning learning rate and number of step and use (lr=0.001, step=10) for MovieLens-20M, (lr=0.001, step=5) for Taobao and (lr=0.001, step=10) for MovieLens-Taobao hybrid dataset.

For the strongest baseline $s^{2}Meta$, we basically adopt the hyper-parameter used in their released codebase \footnote{https://github.com/THUDM/ScenarioMeta}. To basically remove the influence brought by the backbone network, we choose the following backbone network. MovieLens20M: interaction model with [64,32,16] hidden size and ReLU activation function. Taobao: interaction model with [64,64,64] hidden size and ReLU activation function. MovieLen-Taobao: interaction model with [64,64,64] hidden size and ReLU activation function. Note that the interaction model is a model which directly use user-item feature concatenation as input. Other training hyper-parameters are the same with the released code.

Note that it is intractable to conduct completely fair comparison between CMML and $s^{2}Meta$ since they follow completely different inner adaptation strategy. The whole CMML model inherently requires more parameters compared with $s^{2}Meta$ because it is not allowed to conduct gradient descent like $s^{2}Meta$. It has to use additional parameters for context encoder and modulation network to conduct inner-loop adaptation. Even though CMML has more parameters, it still can achieve much higher acceleration compared with $s^{2}Meta$, which can validate the low computational efficiency of gradient based Meta Learning.

For a relatively fair comparison, we ensure the amount of parameters in CMML backbone network is on the same scale compared with that in $s^{2}Meta$. For CMML-Linear, CMML-Sigmoid, and CMML-Film, we use deep network with hidden size [64,64,64] and Leaky ReLU activation function for the backbone network in all three datasets. For CMML-Soft-M, we use 3 layers network and each layer consists of 4 modules. The hidden size of each module is 32. For Soft-M on MovieLens and Taobao, with input size x, the Soft-M has [x*32*4, 32*32*4, 32*64*4] hidden layer and the final [32*1] layer. For Soft-M on MovieLens-Taobao, with input size x, the Soft-M has [x*32*4, 32*32*4, 32*32*4] hidden layer and the final [32*1] layer.  

For LSTM context encoder, we use GRU with hidden size 256 and the output of GRU will also be fed into a MLP with hidden size [256] and ReLU activation function. The output of context encoder has the same dimension with use-item feature concatenation.

For the modulation network, we use a MLP with ReLU activation function. For example, with input size 128 and hidden size 64, CMML-Linear only generates weights for the final weight so it consists of one [128*64, 64*64, 64*output] network. For CMML-Sigmoid/Film/Soft-M, the overall network is [128*64, 64*64, 64*output], with each layer's output extracted to generate the corresponding layer in the backbone network (the output of first 128 * 64 is ReLU activated then goes through one linear layer and get the weight modulation on the first hidden layer of backbone, etc). The hidden size used in our experiment is 128 for Taobao and MovieLens-Taobao and 64 for MovieLens-20M.

And sometimes we find out the bias initialization with small value can increase the stability of the results for CMML-Sigmoid and CMML-FiLM. For these two methods we initialize all MLP's bias with 0.1 in the MovieLens-20M and 0 in Taobao experiment. The other training setting is basically the same with $s^{2} Meta$.
\subsubsection{User-specific setting}
For basaeline DCN and DCN-F, the hyper-parameter setting is the same with that in scenario setting except that learning rate = 1e-3. and we utilise fine-tuning learning rate = 2e-4 and step = 5. And for MELU and MetaDNN baseline, we basically follow the hyperparameter setting of original MELU paper. we modify the learning rate as 3e-4 and it can achieve better result compared with original implementation. For CMML-Film, We use the mean encoder with hidden size [64] and ReLU activation function and backbone network with [64, 64] and Leaky ReLU activation funciton. The modulation network is similar with that of scenario-specific setting except the number of layer becomes 2 and the hidden size is 256. The learning rate is 1e-4 and weight decay is 1e-5.

Our dataset split is different from the original MELU paper because we think the original dataset split violates the basic assumption of Meta Learning - the meta-training distribution and meta-testing distribution should have the same distribution and the split based on time in original MELU paper clearly results in out of distribution problem. Thus we obtain the  dataset by only splitting the user to get warm-warm scenario and warm-cold scenario and retest all baseline on the new dataset. To construct a few-shot setting, we clip the amount of ratings to 100 for users who have more than 100 ratings. The following setting like support-query split is the same with original MELU.

\end{document}